\documentclass[11pt]{article}
\usepackage{epsfig,dsfont,amssymb,amsmath,amsthm,amsfonts,amsbsy,mathrsfs,mathtools}
\usepackage{float}
\usepackage{color}
\usepackage{subfigure}
\graphicspath{{Figures/}}

\usepackage{cite}
\usepackage{comment}
\usepackage{geometry}             
\usepackage{ulem}

\usepackage[hyperindex=true,
          pdfstartview=FitH,
          bookmarksnumbered=true,
          bookmarksopen=true,
          citecolor=blue,
          linkcolor=blue,
          colorlinks=true,
          urlcolor=rossoCP3,
          pdfborder=001,
          unicode]{hyperref}

\definecolor{rossoCP3}{cmyk}{0,.88,.77,.40}

\usepackage{xcolor}  
\definecolor{mygreen}{RGB}{44,85,17}
\definecolor{myblue}{RGB}{34,31,217}
\definecolor{mybrown}{RGB}{194,164,113}
\definecolor{myred}{RGB}{255,66,56}
\definecolor{mypurple}{RGB}{200,36,176}

\allowdisplaybreaks

\begin{document}

\title{Traversability of Schwarzschild-Anti-de Sitter Wormhole \\
in $f(T)$ gravity}

\author{Sheng-Fu Zhang${}^1$, Rui-Hui Lin${}^2$
\vspace{1pt}\\
\small ${}^1$School of Physics, University of Electronic Science and Technology of China, \\ \small No.2006 Xiyuan Ave, West Hi-ech Zone, Chengdu, Sichuan 611731, China\\
\small ${}^2$Division of Mathematical and Theoretical Physics, Shanghai Normal University,\\ \small 100 Guilin Road, Shanghai 200234, China \\
{\small {email}:
\href{202111120514@std.uestc.edu.cn}{202111120514@std.uestc.edu.cn},
\small \href{linrh@shun.edu.cn}{linrh@shun.edu.cn}}}

\date{}
\maketitle

\begin{abstract}
  
In this paper we analyze the traversability of static and evolving Schwarzschild-Anti-de Sitter wormholes. The wormhole metric under consideration is not asymptotically flat. Hence one can only embed this metric into the Euclidean space for a limited radius $r_{max}$. For $r>r_{max}$, an exterior vacuum spacetime should be matched to the wormhole spacetime. In the framework of $f(T)$ gravities, we discuss the null energy condition that the matter supporting the wormhole should satisfy and find that the nontrivial form of $f(T)$ is necessary. For the wormholes to be suitable for human to traverse, we consider the tidal force that a traveler would have felt during his trip. This leads to an upper bound of the traveler's velocity. Utilizing the velocity allowed, we will estimate the travel time through the wormhole. In the evolving cases, the wormhole should not be expanding too fast, otherwise the traveler may not be able to arrive at the other side of the wormhole. Besides this, for static wormholes, we briefly discuss the geodesics in the plane $\theta=\pi/2$.

\end{abstract}


\section{Introduction}

\label{sec:intro}
Wormholes are regarded as hypothetical tunnels connecting two asymptotically flat universes, or two asymptotically portions of the same universe. Consequently, stable wormholes are shortcuts connecting distant regions of spacetime and they can be used for constructing time machines\cite{Cataldo:2017ard}. Primarily, Einstein and Rosen noted that the Schwarzschild black hole has two exterior asymptotic regions connected by a throat. But the Einstein-Rosen wormhole is a space-like and classical object that people cannot pass through it. For our considerable expectation the most physically interesting property of wormholes is that these stable solutions provide a possible, albeit theoretical at the moment, way for time traveling, which, if realized, may lead to the break down of some traditional natural concepts, especially causality.

 After the fundamental paper by Morris and Thorne\cite{Morris1988Wormholes}, the notion of traversable wormholes has gained more and more attention. Morris and Thorne suggested that this stable Lorentzian wormhole is spherically symmetric with a smooth surface and showed that this kind of wormhole might not only allow humans to travel among universe, but also to construct time machines. It is well-known that in classical General Relativity the static spherically symmetric wormholes exist only in the presence of exotic matter, which may have enormous negative energy density and pressure density violating all energy conditions\cite{Cataldo:2017ard,Morris1988Wormholes}, at least in the neighborhood of the wormhole throat. Exotic matter such as phantom matter field\cite{Cataldo:2017yec,Cataldo:2008pm,Cataldo:2008ku} with corresponding energy-momentum tensor $\mathcal{T}_{\mu\nu}$, can be used to support the wormhole geometry. But, the violation of physical reasonableness is not desirable. In their pioneering work\cite{Morris1988Wormholes}, Morris and Thorne also suggested some ways to minimize the effect of exotic matter. Today, due to the observed accelerated expansion of the universe, it seems that there should be a need for such type of matter, that is phantom field as dark energy. But at present, the source of phantom field is yet a mystery. For solving this problem, researchers have studied wormholes in modified theories $f(R)$-gravities\cite{Lobo:2009ip,Saeidi:2011zz,Bhattacharya:2015oma,Francisco2009Wormhole,Pavlovic2015} where $R$ is the Ricci scalar. This is because, particularly, in modified gravities, it may be the effective energy-momentum tensor $\mathcal{T}^{{eff}}_{\mu\nu}$ that violates the energy conditions. The actual matter that threads the wormhole may still satisfy the energy conditions. Similarly, this kind of analysis of wormholes and energy conditions also has been performed in various alternatives of gravitation theories, e.g., $f(T)$ gravities\cite{Jamil:2012ti}, $T$ the torsion scalar, Einstein-Gauss-Bonnet theory\cite{Mehdizadeh:2015jra}, and Einstein-Cartan gravity\cite{Mohammad2017Dynamic,MehdizadehEinstein},etc..If by chance, we consider wormhole models within the framework of modified gravitation theories provided above or other possibilities, the presence of exotic matter may be not necessary\cite{Harko:2013yb,MoraesNon}. For higher dimensional cases, generally 5 dimensions\cite{NajafiFive} or more, the Kaluza-Klein theory\cite{article2} can be utilized for wormhole investigations, and more especially, N-dimensional wormholes\cite{Zangeneh:2014noa,Mauricio,Mauricio2011N} also have been constructed with considerable achievements.

Note that the Schwarzchild and Schwarzchild-Anti-de Sitter (S-AdS) solutions can be interpreted as wormholes, which, however, are not suitable for traverse due to the horizon at the throat. Researchers can thus construct wormholes with the same spatial shape, but they require that the redshift function does not lead to a horizon\cite{Morris1988Wormholes}. This kind of wormholes, like other GR solutions, can be admitted in $f(R)$ gravities and $f(T)$ gravities for a certain choice of the proper tetrad\cite{Lin2019}. They can also resolve the energy condition problems by leaving them to the modification of the gravitation theory. Besides the static wormholes, the wormholes in the expanding universe may also evolve over time\cite{Cataldo:2008pm,Cataldo:2008ku,Saeidi:2011zz,Bhattacharya:2015oma}. Moreover, once the energy conditions are checked, these wormholes can then submit to other criteria for practical spacetime traverse, such as the tidal force and gravitational acceleration the travelers feel, or the total travel time required\cite{Cataldo:2017ard}, which were also pointed out by Morris and Thorne\cite{Morris1988Wormholes} in their pioneering work, but are much less explored in other literature concerning the traversability of wormholes.

In this work, we will pay much attention to the static and evolving S-AdS wormhole solutions. We consider the geometry and the NEC for these wormholes in the $f(T)$ gravity framework. We also look into the tidal force and the total travel time that an explorer may practically feel about, and hence discuss the criteria of the traversability of the S-AdS wormholes. This paper is organized as follows. In Sec.2, we briefly review the wormhole geometry and show the chosen proper tetrad, and then we obtain the S-AdS wormhole solutions from the equation of motion and analyze the null energy condition (NEC) in $f(T)$ gravity framework. We investigate the traversability and geodesics of the static S-AdS wormholes in Sec.3. Section 4 gives the analyses for the evolving S-AdS wormholes with two concrete and different types of scale factors $R(t)$, respectively. Section 5 contains our main conclusions and discussions. Particularly, it should be emphasized that we will be using the unit $c=G=1$ throughout this paper.

\section{Wormhole Geometry and Null Energy Condition}
\label{sec:2}

\subsection{Wormhole Geometry}
First, let us consider a spherically symmetric wormhole that is isotropically evolving
\begin{equation}
\begin{split}\label{Wmetric}
ds^2=e^{2\psi(r)}dt^2-R(t)^{2}\Big(\frac{1}{1-\frac{b(r)}{r}}dr^2+r^2d\theta^2+r^2\sin^2 \theta d\phi^2\Big),
\end{split}
\end{equation}
where $\psi(r)$ and $b(r)$ denote the redshift function and the shape function of the wormhole, respectively, and $R(t)>0$ is the scale factor.

For the wormhole to be traversable, the background spacetime described by Eq.\eqref{Wmetric} should not lead to any horizon and singularity. Thus the redshift function $\psi(r)$ must be finite everywhere. Specifically, we set $\psi(r)$ to be a constant for simplicity, and then it can be absorbed into the temporal coordinate by a universal rescale of time. Note that the metric \eqref{Wmetric} only describes the spacetime outside the throat of the wormhole $r\ge r_{0}$, where $r_{0}$ represents the coordinate position of the throat. For the wormhole to be actually flaring out, the shape function $b(r)$ should satisfy\cite{Morris1988Wormholes}
\begin{gather}\label{focondition}
b(r_{0})=r_{0},\quad b'(r_0)<1,
\end{gather}
in the neighborhood of the throat for any given time. For metric \eqref{Wmetric}, to describe a physical wormhole, $b(r)$ also needs to satisfy
\begin{gather}\label{pcondition}
b(r)\ge0, \quad 1-\frac{b(r)}{r}\ge0,
\end{gather}
such that the radial coordinate does not change signs. Moreover, it should be noted that the metric Eq.\eqref{Wmetric} may be not asymptotically flat. From an embedding point of view, the wormholes described by Eq.\eqref{Wmetric} can only have an effective and finite size with a maximum radius $r_{max}$. Beyond $r_{max}$, one may need to match this wormhole metric Eq.\eqref{Wmetric} to an exterior vacuum spacetime\cite{Garcia2011,Rosa2018}

For simplicity we will set $\psi(r)=1$ and find that the following choice of proper tetrad ${e^a}_\mu$\cite{Lin2019} for the metric \eqref{Wmetric} can be adopted,
\begin{align}\label{tetrad}
e^{\hat{0}}&=dt, \notag \\
e^{\hat{1}}&= \begin{multlined}[t]
R(t)\sqrt{\frac{1}{1-\frac{b(r)}{r}}}\sin\theta\cos\phi dr \notag \\  
+R(t)\Big(\sqrt{1-\frac{b(r)}{r}}\cos\theta\cos\phi\pm\sqrt{\frac{b(r)}{r}}\sin\phi \Big)rd\theta \\
-R(t)\Big(\sqrt{1-\frac{b(r)}{r}}\sin\phi\mp\sqrt{\frac{b(r)}{r}}\cos\theta\cos\phi \Big)r\sin\theta d\phi,
\end{multlined} \\
e^{\hat{2}}&=\begin{multlined}[t]
R(t)\sqrt{\frac{1}{1-\frac{b(r)}{r}}}\sin\theta\sin\phi dr \\
+R(t)\Big(\sqrt{1-\frac{b(r)}{r}}\cos\theta\sin\phi\mp\sqrt{\frac{b(r)}{r}}\cos\phi \Big)rd\theta  \\
+R(t)\Big(\sqrt{1-\frac{b(r)}{r}}\cos\phi\pm\sqrt{\frac{b(r)}{r}}\cos\theta\sin\phi \Big)r\sin\theta d\phi,
\end{multlined} \\
e^{\hat{3}}&=\begin{multlined}[t]
R(t)\sqrt{\frac{1}{1-\frac{b(r)}{r}}}\cos\theta dr \notag \\
-R(t)r\sqrt{1-\frac{b(r)}{r}}\sin\theta d\theta\mp R(t)r\sqrt{\frac{b(r)}{r}}\sin^{2}\theta d\phi.
\end{multlined}
\end{align}
Then the above metric \eqref{Wmetric} and proper tetrad \eqref{tetrad} can lead to the corresponding torsion scalar
\begin{gather}
T(r,t)=\frac{2b'}{r^2R^2}-\frac{6\dot{R}^2}{R^2},
\end{gather}
where the prime and dot symbols denote the derivatives with respect to coordinates $r$ and $t$, respectively.

\subsection{Wormhole Solution and Energy Condition}

Here we consider wormholes in the framework of $f(T)$ gravities given by the following action
\begin{gather}
S=\frac{1}{16\pi }\int ef(T)d^4x+\int e\mathcal{L}_md^4x,
\end{gather}
where $T$ is the torsion scalar, $\mathcal{L}_{m}$ is the Lagrangian density due to the matter fields, $e$ is the determinant of the proper tetrad and $f(T)$ is an arbitrary function of $T$. It is noted that for special case $f(T)=-T$, the $f(T)$-gravity becomes back to the Teleparallel Equivalent of General Relativity (TEGR). Then take the variation of the action with respect to the proper tetrad, we will obtain the equation of motion
\begin{equation}\label{eom}
4({e^a}_\nu\partial_\mu(e{S_a}^{\mu\sigma})-e{T^\rho}_{\mu\nu}{S_\rho}^{\sigma\mu})f_T+4e(\partial_\mu T){S_\nu}^{\mu\sigma}f_{TT}-e{\delta^\sigma}_\nu f(T)=16\pi e{\mathcal{T}^\sigma}_\nu,
\end{equation}
where ${T^\rho}_{\mu\nu}$ is the torsion tensor and ${S_\rho}^{\mu\nu}$ is the superpotential, ${\mathcal{T}^\mu}_\nu$ denotes the energy-momentum tensor for the matter fields, $f_T$ and $f_{TT}$ have definitions as $\partial f(T)/\partial T$ and $\partial^2f(T)/\partial T^2$, respectively.

For an anisotropic fluid which threads the wormhole, it is given by\cite{Cataldo:2017ard,Morris1988Wormholes}
\begin{equation}
\begin{split}
{\mathcal{T}^\sigma}_\nu=(\rho+P_t)u^\sigma u_\nu-P_t{\delta^\sigma}_\nu+(P_r-P_t)v^\sigma v_\nu,
\end{split}
\end{equation}
where $u^{\mu}$ is the 4-velocity vector, $v^{\mu}$ is the unitary space-like vector in the radial direction, $\rho$ is the energy density, and $P_{r}$,$P_{t}$ are the radial and tangent pressure, respectively. Utilizing this energy-momentum tensor with the equation of motion \eqref{eom}, the non-diagonal components read 
\begin{align}
r^2R^2\dot{R}\sin\theta (\partial_rT)f_{TT}&=0, \notag \\
rR\sqrt{\frac{b}{r}}(\partial_rT)f_{TT}&=0,
\end{align}
which immediately have the analytical solutions 
\begin{gather}\label{solution}
b(r)=\frac{\Lambda_0}{3}r^3-m,
\end{gather}
where $\Lambda_0$ and $m$ are constants. From the previous condition \eqref{focondition} we have
\begin{gather}
\Lambda_0=\frac{3}{r_0^3}m+\frac{3}{r_0^2}, \quad \quad m<-\frac{2}{3}r_0,
\end{gather}
and therefore, $m$ is a negative constant. Because we want the shape function given by Eq.\eqref{solution} to describe an S-AdS wormhole, we assume $\Lambda_0<0$ hereinafter, and then we immediately obtain $m<-r_{0}$. Note that for $m<-r_{0}$, it is not difficult to find that $1-b/r$ is positive only in a finite neighborhood of the throat, resulting in the existence of a maximum radius $r_{max}$ such that once beyond $r_{max}$, the condition \eqref{pcondition} will be violated. The maximum radius can be determined by
\begin{equation}
 r_{max}=\sqrt[3]{\frac{m}{m+r_0}}r_0.
\end{equation}
Finally we conclude for S-AdS wormholes that
\begin{gather}\label{gcondition}
r_{0}\le r\le r_{max}, \quad \Lambda_0=\frac{3}{r_0^3}m+\frac{3}{r_0^2}, \quad  m<-r_0.
\end{gather}

The diagonal parts of the field equations read
\begin{align}\label{fequations}
16\pi\rho&=-\frac{12\dot{R}^2}{R^2}f_T-f(T), \notag \\
16\pi P_r&=\frac{12\dot{R}^2}{R^2}f_T+f(T)-2\Big(\frac{2\Lambda_0}{3R^2}+\frac{m}{r^3R^2}\Big)f_T+4\partial_t\Big(\frac{\dot{R}}{R}f_T \Big), \\
16\pi P_t&=\frac{12\dot{R}^2}{R^2}f_T+f(T)-2\Big(\frac{2\Lambda_0}{3R^2}-\frac{m}{2r^3R^2}\Big)f_T+4\partial_t\Big(\frac{\dot{R}}{R}f_T\Big), \notag
\end{align}
where Eq.\eqref{solution} has been used. The NEC $\mathcal{T}_{\mu\nu}k^{\mu}k^{\nu}\ge0$, where $k^{\mu}$ is any null vector, implies that
\begin{equation}
\rho+P_r\ge0 ,\quad \quad \rho+P_t\ge0,
\end{equation}
and then together with Eq.\eqref{fequations}, it follows that
\begin{align}\label{necondition}
2\partial_t\Big(\frac{\dot{R}}{R}f_T\Big)&\ge\Big(\frac{2\Lambda_0}{3R^2}+\frac{m}{r^3R^2}\Big)f_T \notag \\
2\partial_t\Big(\frac{\dot{R}}{R}f_T\Big)&\ge\Big(\frac{2\Lambda_0}{3R^2}-\frac{m}{2r^3R^2}\Big)f_T.
\end{align}

So far we have considered generally the geometry of the S-AdS wormholes and the null energy condition. For the wormholes to be practically traversable, we also need to pay our attention to what a human traveler would have gone through while traversing the wormhole. In the following sections, we will discuss the tidal force, traveling speed, and total travel time, as well as the NEC in some concrete cases.

\section{Static S-AdS Wormholes}
\label{sec:3}

Firstly, we investigate the static case with $R(t)=1$. It is easy to find that, together with the condition \eqref{gcondition}, the NEC can be satisfied by
\begin{equation}
f_{T}\ge0, \quad  \Lambda_0\le-\frac{1}{r_0^2}, \quad  m\le-\frac{4}{3}r_0.
\end{equation}
At this point, we can note that in the Teleparallel Equivalent of General Relativity, namely $f(T)=-T$, S-AdS wormholes cannot be supported by normal matter which satisfies the NEC. The presence of exotic matter is consistent with the conclusions in GR\cite{Cataldo:2017ard,Morris1988Wormholes}, and therefore nontrivial form of $f(T)$ is necessary for non-exotic matter to thread this wormhole.

\subsection{Traversability for Static Wormhole}

With the wormhole supported by $f(T)$ framework and normal matter, we then can consider the traverse of the wormhole. Note that one of the criteria of traversability is the gravitational tidal force felt by the traveler. Different parts of the traveler, due to the extent of his body, will experience different accelerations in sufficiently strong gravity, leading to stretching or squeezing of his body. Suppose that the traveler always keeps in the equator $\theta=\pi/2$ plane and traverses radially through the wormhole, beginning at space station 1 with coordinate position $r_{1}$ on one side of wormhole throat, and ending at space station 2 with coordinate position $r_{2}$ on the other side. The traversability condition requires that the tidal acceleration should not exceed the Earth's gravitational acceleration, namely $\left|\Delta\boldsymbol a\right|\le g_\oplus$, which directly gives the radial and lateral tidal conditions, respectively\cite{Cataldo:2017ard,Morris1988Wormholes}
\begin{align}\label{tcondition}
\left|(1-\frac{b}{r})(\psi''-\frac{rb'-b}{2r(r-b)}\psi'-(\psi')^2)\right|\left|\xi\right|&\le g_\oplus, \notag \\
\left|\frac{\gamma^2}{2r^2}[v^2(b'-\frac{b}{r})+2(r-b)\psi']\right|\left|\xi\right|&\le g_\oplus,
\end{align}
where $\left|\xi\right|$ denotes the size of the traveler's body, $v$ is the traveler's velocity, $\gamma=(1-v^2)^{-1/2}$ is the Lorentz factor, $g_\oplus$ is the Earth's gravitational acceleration. Note that the radial tidal condition should be regarded as the constraint on the redshift function $\psi$, which has been directly satisfied by setting $\psi(r)=1$ everywhere, while the lateral condition should be regarded as constraint on the traveler's velocity $v$ throughout his journey. Substituting Eq.\eqref{solution} into Eq.\eqref{tcondition} leads to
\begin{equation}
\left|\frac{v^2}{2r^3}(\frac{2}{3}\Lambda_0r^3+m)\right|\left|\xi\right|\le g_\oplus,
\end{equation}
where we have utilized the non-relativistic limit, namely $v<<c$ and $\gamma\approx1$. It is obviously that this inequality gives an upper bound of the traveler's velocity in his trip, which is more stringent at the smaller radius, hence evaluating at the throat $r_{0}$ we obtain
\begin{equation}
v\le r_{0}\sqrt{\frac{2g_\oplus r_{0}}{\left|3m+2r_0\right|\left|\xi\right|}}=r_{0}\sqrt{\frac{2g_\oplus}{\left|3\chi+2\right|\left|\xi\right|}},
\end{equation}
where we have introduced a new argument $\chi=m/r_{0}$. Using the NEC $\chi\le-\frac{4}{3}$, we immediately have
\begin{equation}
v\le r_{0}\sqrt{\frac{2g_\oplus}{\left|3\chi+2\right|\left|\xi\right|}}\le r_{0}\sqrt{\frac{g_\oplus}{\left|\xi\right|}}.
\end{equation}
This gives an upper bound of the constant traveling velocity, which then gives rise to the concern about the total travel time.

As shown in Ref.\cite{Morris1988Wormholes}, the proper distance in the wormhole spacetime is defined by
\begin{equation}\label{pdistance}
dl=\frac{dr}{\sqrt{1-\frac{b(r)}{r}}},
\end{equation}
also the coordinate time lapse $dt$ and the proper time lapse $d\tau$ of the traveler are
\begin{equation}
dt=\frac{dl}{v}, \quad d\tau=\frac{dl}{\gamma v}\simeq dt,
\end{equation}
we can find that in the non-relativistic motion, the proper time $\tau$ ticked by clocks is equal to the coordinate time $t$. If we want to consider the trip starting from and ending at space stations outside of the wormhole with $r_{1},r_{2}>r_{max}$, the whole travel time then should be $t_{ext}+\Delta t$, where $t_{ext}$ denotes the time spent in the exterior vacuum space before reaching and after leaving $r_{max}$, and $\Delta t$ is the needed to traverse the wormhole within $r_{max}$. Because the trip outside $r_{max}$ is in the ordinary vacuum, so we only concern about the reasonable part $\Delta t$ spent within $r_{max}$ in the following calculations.

For S-AdS wormholes in this paper, the total proper distance from the position $r=kr_{0}$ with $k>1$ to the throat $r_{0}$ is given by
\begin{equation}
\begin{split}
\Delta l(k)&=r_{0}\int_{1}^{k}\frac{\sqrt{s}ds}{\sqrt{-(\chi+1)s^3+s+\chi}}, \quad 1<k\le k_{max}= \sqrt[3]{\frac{\chi}{\chi+1}},
\end{split}
\end{equation}
and for simplicity in the following discussion, we denote the integral in $\Delta l$ as
\begin{equation}\label{Gfunction}
\mathcal{G}(k,\chi)=\int_{1}^{k}\frac{\sqrt{s}ds}{\sqrt{-(\chi+1)s^3+s+\chi}}.
\end{equation}
Now it is easy to compute the total travel time spent between the station 1 and station 2 with $r_{1,2}=kr_{0}$, and the result shows
\begin{equation}
\Delta\tau\simeq\Delta t=\frac{2\Delta l}{v}=\frac{2r_{0}}{v}\mathcal{G}(k,\chi).
\end{equation}
Suppose that the space stations 1 and 2 are both located at $r_{1,2}=r_{max}=k_{max}r_{0}$ and the traveler always takes largest velocity allowed, then the total travel time is 
\begin{equation}
\Delta t(\chi)=2\sqrt{\frac{\left|3\chi+2\right|\left|\xi\right|}{2g_\oplus}}\mathcal{G}\Big(\sqrt[3]{\frac{\chi}{\chi+1}} ,\chi\Big),
\end{equation}
which is a monotonically increasing function of $\chi$. It must be emphasized that this manipulation is our most conservative estimation of the traverse time. Utilizing $g_\oplus\simeq9.8m/s^2$ and taking $\xi\simeq2m$, we can then calculate that
\begin{equation}
0s<\Delta t\le 1.014s, \quad  -\infty<\chi\le-\frac{4}{3}.
\end{equation}

\subsection{Geodesics}

Before discussion, we must point out that the geodesics described in this subsection are always located within $r_{max}$. The Ref.\cite{Cataldo:2017ard} has told us that the geodesic behavior in the wormhole spacetime can be obtained by analyzing the Lagrangian $\mathcal{L}=\frac{1}{2}g_{\mu\nu}\frac{dx^\mu}{d\tau}\frac{dx^\nu}{d\tau}$, namely
\begin{equation}
\mathcal{L}=\frac{1}{2}(\dot{t}^2-\frac{\dot{r}^2}{1-\frac{\Lambda_0}{3}r^2+\frac{m}{r}}-r^2\dot{\theta}^2-r^2\sin^2\theta\dot{\phi}^2),
\end{equation}
in this paper for S-AdS wormholes, where $\tau$ is the proper time and the dot denotes the derivative with respect to $\tau$ only in this subsection. And then the conserved conjugate momenta $\Pi_t$ and $\Pi_\phi$ are given by 
\begin{align}
\Pi_t&=+\frac{\partial\mathcal{L}}{\partial\dot{t}}=\dot{t}=E,\\
\Pi_\phi&=-\frac{\partial\mathcal{L}}{\partial\dot{\phi}}=r^2\sin^2\theta\dot{\phi}.
\end{align}
Generally, we only consider the test particle paths located at the plane $\theta=\pi/2$, and therefore we have
\begin{equation}\label{gequation}
E^2=h+\frac{L^2}{r^2}+\frac{\dot{r}^2}{1-\frac{\Lambda_0}{3}r^2+\frac{m}{r}}, \quad r^2\dot{\phi}=L,
\end{equation}
in which $h=1$ for time-like geodesics and $h=0$ for null geodesics, and $L$ is the angular momentum per unit mass.

For radial time-like geodesics with $L=0$ and $h=1$, specifically, the initial position $(\tau_{i},r_{i})$ and initial velocity mean that the test particle, at $\tau_{i}=0$, starts to move at $r_{i}$ with initial velocity $v_{i}$, then we have
\begin{equation}\label{rgequation}
E^2=1+\frac{v_i^2}{1-\frac{\Lambda_0}{3}r_i^2+\frac{m}{r_i}},
\end{equation}
and immediately Eq.\eqref{gequation} can be changed into
\begin{equation}\label{rgequation2}
\dot{r}=\pm\sqrt{\frac{1-\frac{\Lambda_0}{3}r^2+\frac{m}{r}}{1-\frac{\Lambda_0}{3}r_i^2+\frac{m}{r_i}}v_i^2},
\end{equation}
where the $-$ sign is valid for particles moving from $r_{i}$ towards the wormhole throat, while the $+$ sign for particles moving away from $r_{0}$. Note that if the particle has zero initial velocity, then $\dot{r}=0$ and $r(\tau)=const$, which implies that the test particle will remain at rest at the initial position $r_{i}$. In other words, we need to give the test particle an initial velocity $v_{i}\ne0$ in order to push it towards the wormhole throat or the opposite direction. Additionally, it is very easy to note that the test particle must have zero radial velocity at the throat. On the other side, Eq.\eqref{rgequation} tells us that
\begin{equation}
v_i^2=(E^2-1)(1-\frac{\Lambda_0}{3}r_i^2+\frac{m}{r_i}).
\end{equation}
Then we conclude that $E^{2}\ge1$ due to the geometry conditions of S-AdS wormhole. By deriving $\dot{r}$ with respect to the affine parameter $\tau$, we obtain the second derivative of $r$ written as 
\begin{align}
\ddot{r}=-\frac{1}{2}(E^2-1)(\frac{2}{3}\Lambda_0r+\frac{m}{r^2}),
\end{align} 
which is positive in the wormhole spacetime, showing us that the gravitational field is repulsive everywhere. Together with the definition \eqref{pdistance}, the equation of motion \eqref{rgequation2} also leads to $\dot{l}^2=E^2-1$ and $\ddot{l}=0$. Then for radially moving particle, the condition $E^2=1$ implies that the test particle is at rest with a zero radial acceleration.

For the non-radial timelike geodesics we set $L\ne0$ and $h=1$. The relevant initial conditions are considered as the initial position $(r_{i},\phi_{i}=0)$ and the initial radial velocity $v_{i}$ and the initial angular velocity $\dot{\phi}_{i}$. This implies that at $\tau_{i}$, the test particle starts to move at $r_{i}$ with initial radial and angular velocity $v_{i}$ and $\dot{\phi}_{i}$, respectively. Therefore Eq.\eqref{gequation} reads
\begin{equation}
E^2=1+\frac{L^2}{r_i^2}+\frac{v_i^2}{1-\frac{\Lambda_0}{3}r_i^2+\frac{m}{r_i}},
\end{equation}
from which we conclude that $E^2>1$. Also note that $L=r^2\dot{\phi}=r_{i}^2\dot{\phi}_{i}$, then the equation of motion can be written as
\begin{align}\label{nrgequation}
\dot{r}^2= (1-\frac{\Lambda_0}{3}r^2+\frac{m}{r})\Big(L^2(\frac{1}{r_i^2}-\frac{1}{r^2})+\frac{v_i^2}{1-\frac{\Lambda_0}{3}r_i^2+\frac{m}{r_i}}\Big),
\end{align}
and radial acceleration is obtained by
\begin{align}\label{racceleration}
\ddot{r}=-\frac{1}{2}(\frac{2}{3}\Lambda_0r+\frac{m}{r^2})\Big(L^2(\frac{1}{r_i^2}-\frac{1}{r^2})+\frac{v_i^2}{1-\frac{\Lambda_0}{3}r_i^2+\frac{m}{r_i}}\Big)+(1-\frac{\Lambda_0}{3}r^2+\frac{m}{r})\frac{L^2}{r^3}.
\end{align}

Note that for circular geodesics with $r>r_{0}$, its radial velocity and radial acceleration must vanish, but the zero radial velocity $\dot{r}=0$ will cause Eq.\eqref{racceleration} to become as
\begin{equation}
\ddot{r}=(1-\frac{\Lambda_0}{3}r^2+\frac{m}{r})\frac{L^2}{r^3},
\end{equation}
which is obviously positive in this spacetime. For circular geodesics with $r=r_{0}$, the vanished radial velocity and acceleration can be satisfied by 
\begin{equation}
r=r_0=\frac{\left|L\right|}{\sqrt{E^2-1}}.
\end{equation}
These analyses directly point out that the circular orbits only exist at the wormhole throat. In other words, if the test particle has zero radial velocity at any other position $r>r_{0}$, then it must be always accelerated in the direction of increasing $r$.

For non-circular geodesics, the equation of motion \eqref{nrgequation} tells us that the test particle has zero radial velocity at $r_{0}$ and at the radius
\begin{equation}
r_m=\Big(\frac{1}{r_i^2}+\frac{v_i^2}{L^2(1-\frac{\Lambda_0}{3}r_i^2+\frac{m}{r_i})}\Big)^{-\frac{1}{2}}=\frac{\left|L\right|}{\sqrt{E^2-1}}<r_i.
\end{equation}
This result implies that if $r_{m}>r_{0}$, then the radial position $r_{m}$ represents a reversal point for a geodesic starting at $r_{i}$ with $v_{i}<0$ and $\dot{\phi}_{i}\ne0$. Therefore the test particle, which approaches to the throat, will be reflected at this radius $r_{m}$ by the wormhole. In other words, $r_{m}$ is the minimum distance to the wormhole throat which can be reached by the test particle. In this case, the timelike geodesic lies always on the one side of the wormhole and the test particle cannot pass through the throat.
\begin{figure}[htbp]
	\centering
	\subfigure[$m=-2$]
	{\begin{minipage}{6cm}
			\centering
			\includegraphics[scale=0.56]{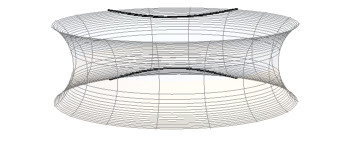}
	\end{minipage}}
	\subfigure[$m=-3/2$]
	{\begin{minipage}{6cm}
			\centering
			\includegraphics[scale=0.40]{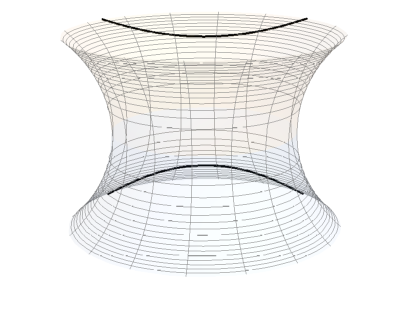}
	\end{minipage}}
	\caption{Plots show that non-radial geodesics with $r_0=1$, $E^2=4$ and $L^2=9$ do not cross the wormhole throat.}
	\label{Fig.1}
\end{figure}

Note that the reversal points do exist only for $r_{m}\ge r_{0}$, namely
\begin{equation}\label{constraint}
 \frac{L^2}{(E^2-1)}\ge r_{0}^2.
\end{equation}
On the other hand, Eq.\eqref{nrgequation}, together with the radial proper distance \eqref{pdistance}, makes us obtain $\dot{l}^2=E^2-1-\frac{L^2}{r^2}$, where $\dot{l}=\frac{dl}{d\tau}$. The turning points require $\dot{l}=0$, which is easily solved by $r_{m}=\frac{\left|L\right|}{\sqrt{E^2-1}}$, being consistent with Eq.\eqref{constraint}. Relatively, if these constants $E$, $L$ and $r_{0}$ do not satisfy the condition \eqref{constraint}, then the reversal points cannot exist and the non-radial geodesics will pass to the other side of the throat.
\begin{figure}[htbp]
	\centering
	\subfigure[$m=-2$]
	{\begin{minipage}{6cm}
			\centering
			\includegraphics[scale=0.5]{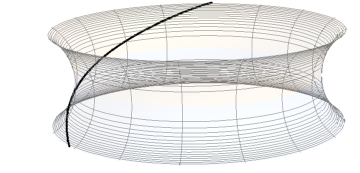}
	\end{minipage}}
	\subfigure[$m=-3/2$]
	{\begin{minipage}{6cm}
			\centering
			\includegraphics[scale=0.41]{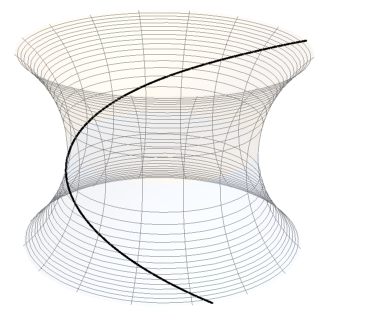}
	\end{minipage}}
	\caption{Plots show that non-radial geodesics with $r_0=1$, $E^2=4$ and $L^2=1$ pass through the wormhole throat.}
	\label{Fig.2}
\end{figure}

\section{Evolving S-AdS Wormholes}
\label{sec:4}

For more complex evolving cases, we only propose some intriguing viewpoints and investigate the S-AdS wormholes roughly in this section, and this implies that we may lose some important details making our results incorrect. Because all of what we will demonstrate below are based on some simple assumptions, those results obtained by our calculations may not happen in the real universe.

Similar to the static cases, for a radial traveler, the radial and lateral tidal accelerations are required to satisfy\cite{Cataldo:2008pm}
\begin{align}\label{etcondition}
\left|\frac{\ddot{R}}{R}\right|\left|\xi\right|&\le g_\oplus, \notag \\
\left|\frac{\gamma^2\ddot{R}}{R}-\frac{\gamma^2v^2}{2r^3R^2}\Big(2r^3\dot{R}^2+rb'-b\Big)\right|\left|\xi\right|&\le g_\oplus.
\end{align}
Substituting Eq.\eqref{solution} into Eq.\eqref{etcondition} gives
\begin{align}\label{etcondition2}
\left|\frac{\ddot{R}}{R}\right|\left|\xi\right|&\le g_\oplus, \notag \\
\left|\frac{\ddot{R}}{R}-\frac{v^2}{2R^2}(\frac{m}{r^3}+\frac{2}{3}\Lambda_0+2\dot{R}^2)\right|\left|\xi\right|&\le g_\oplus,
\end{align}
where we have utilized $\gamma\simeq1$ for the nonr-elativistic limit

In the evolving case, the radial proper distance from position $r=kr_{0}$ with $k>1$ to the throat $r_{0}$ is defined by
\begin{equation}
\begin{split}
l(r,t)= R(t)\int_{r_0}^{r}\frac{dr}{\sqrt{1-\frac{\Lambda_0}{3}r^2+\frac{m}{r}}}=r_{0}R(t)\mathcal{G}(k,\chi),
\end{split}
\end{equation}
where $\mathcal{G}(k,\chi)$ is given by Eq.\eqref{Gfunction}. The velocity of the traveler then satisfies
\begin{equation}
\begin{split}\label{eomotion}
v=\frac{dl}{dt}=r_{0}(\dot{R}\mathcal{G}+R\dot{\mathcal{G}}).
\end{split}
\end{equation}
Hence if the traveler takes constant velocity $v=\pm r_{0}\nu$ in his trip, where $\nu$ is also a constant and the $+$ and $-$ signs correspond to the outward and inward movements, respectively, then in the first part of his trip beginning at the space station 1 with $r_{1}=k_{1}r_{0}$ and going towards the throat $r_{0}$, Eq.\eqref{eomotion} with $v=-r_{0}\nu$ gives 
\begin{equation}
R(t)\mathcal{G}(k,\chi)=\mathcal{G}(k_{1},\chi)-\nu t.
\end{equation}
Suppose $R(0)=1$ and note that at the throat, $\mathcal{G}(1,\chi)=0$, we can easily find the moment $t_{0}$ when the traveler arrives in the throat satisfying
\begin{equation}
t_{0}=\frac{\mathcal{G}(k_{1},\chi)}{\nu}.
\end{equation}
Similarly, for the second part of his trip that the traveler leaves the throat at $t_{0}$, Eq.\eqref{eomotion} with $v=r_{0}\nu$ equals solution
\begin{equation}
R(t)\mathcal{G}(k,\chi)=\nu(t-t_{0}).
\end{equation}
Therefore, when the traveler reaches the space station2 at $r_{2}=k_{2}r_{0}$, the total travel time $t$ can be given by
\begin{equation}\label{tsolution}
\nu t=\mathcal{G}(k_{2},\chi)R(t)+\mathcal{G}(k_{1},\chi).
\end{equation}
But if the traveler takes the accelerated motion with velocity $v=\pm r_{0}\nu R(t)$ in his trip, the similar analysis gives the total traverse time $t$ satisfying
\begin{equation}\label{tsolution2}
\nu\int_{0}^{t}Rdt=\mathcal{G}(k_{2},\chi)R(t)+\mathcal{G}(k_{1},\chi).
\end{equation}
Since we only concern about the part of the spacetime within $r_{max}$, for simplicity, we set the station positions at $r_{1}=r_{2}=k_{max}r_{0}$ with $k_{max}=(\chi/(\chi+1))^{1/3}$ in the following discussion.

Now we consider the \textbf{Model.A} with a linear scale factor $R(t)=\alpha t+1$, here $\alpha$ being a positive parameter indicating the expanding rate of the wormhole. Substituting this scale factor into Eq.\eqref{necondition}, then the NEC can be satisfied by
\begin{equation}
f_{T}\ge0, \quad f_{TT}\ge0, \quad  \chi\le-\frac{4}{3}(1+4\alpha^2r_{0}^2).
\end{equation}
Note that the radial tidal force condition in Eq.\eqref{etcondition2} has been naturally satisfied by the present case. However for the lateral constraint, we find that it is more stringent at smaller radius, and therefore, at the throat we obtain
\begin{equation}\label{vcondition}
v\le r_{0}R\sqrt{\frac{2g_\oplus}{\left|3\chi+2+2\alpha^2r_{0}^2\right|\left|\xi\right|}},
\end{equation}
where we have used $m=\chi r_{0}$. Similarly, this upper bound of the velocity gives rise to the concern about travel time. One can see the allowable maximum velocity depending on $R(t)$. If the traveler wants to go through the wormhole as fast as possible, then he should adjust his velocity over time using Eq.\eqref{vcondition}. In the following consideration of the total travel time, we assume that the traveler always travels at the largest velocity allowed.

Now Eq.\eqref{vcondition} decides the notation $\nu_{A}$ in the traveler's speed $v$ with
\begin{equation}
\nu=\nu_{A}\equiv\sqrt{\frac{2g_\oplus}{\left|3\chi+2+2\alpha^2r_{0}^2\right|\left|\xi\right|}}.
\end{equation}
In the present case, if the traveler takes constant velocity, namely $v=r_{0}\nu_{A}$, the total travel time spent in his trip between station1 and station2 can be obtained from Eq.\eqref{tsolution} as
\begin{equation}
\Delta t(\chi)=\frac{2\mathcal{G}(\sqrt[3]{\frac{\chi}{1+\chi}},\chi)}{\nu_{A}-\alpha\mathcal{G}(\sqrt[3]{\frac{\chi}{1+\chi}},\chi)}.
\end{equation}
Immediately, we must require
\begin{equation}
\nu_{A}-\alpha\mathcal{G}(\sqrt[3]{\frac{\chi}{1+\chi}},\chi)>0.
\end{equation}
For example, if $\chi=-\frac{4}{3}(1+4\alpha^2r_{0}^2)$ and there exists a wormhole having a radius of throat $r_{0}=10^5m$ and expanding rate $\alpha=5\times10^{-6}/s$, then the velocity is $v\simeq0.00044c\simeq133485m/s$ and it will take the traveler $\Delta t=0.505634s$ to traverse the wormhole. On the other hand, if the traveler's velocity is $v=r_{0}R\nu_{A}$, Eq.\eqref{tsolution2} can give the total travel time as
\begin{equation}
\Delta t(\chi)=\frac{2}{\nu_{A}}\mathcal{G}(\sqrt[3]{\frac{\chi}{1+\chi}},\chi),
\end{equation}
and then the traveler will spend $\Delta t=0.505633s$ to finish his trip. Here,  because of the non-relativistic limit, then it is necessary to check that he will not be too fast when he arrives at the destination. It is easy to find that his final velocity will be $v_{f}\simeq0.00044c$, so the result which we obtained is valid.

For a second evolving wormhole \textbf{Model.B} with an exponential scale factor $R(t)=e^{\beta t}$, where $\beta$ is a positive parameter, the NEC \eqref{necondition} gives
\begin{equation}
f_{T}\ge0, \quad f_{TT}\ge0, \quad \chi\le-\frac{4}{3},
\end{equation}
where we have utilized $\chi=m/r_{0}$. From Eq.\eqref{etcondition2}, the radial constraint of tidal force directly reads $\beta^2\left|\xi\right|\ge g_\oplus$, while at the throat, the lateral restriction gives
\begin{equation}
v\le r_{0}R\sqrt{\frac{2(g_\oplus-\left|\xi\right|\beta^2)}{\left|3\chi+2\right|\left|\xi\right|}}.
\end{equation}
Similar to the first model, we can find the notation $\nu_{B}$ in this case
\begin{equation}
\nu_{B}\equiv\sqrt{\frac{2(g_\oplus-\left|\xi\right|\beta^2)}{\left|3\chi+2\right|\left|\xi\right|}}.
\end{equation}

Hence if the traveler takes constant speed $v=r_{0}\nu_{B}$ in his trip from station1 to station2, Eq.\eqref{tsolution} tells us that the total travel time $t$ should satisfy
\begin{equation}
\frac{t}{e^{\beta t}+1}=\frac{\mathcal{G}(\sqrt[3]{\frac{\chi}{1+\chi}},\chi)}{\nu_{B}}.
\end{equation}
Taking $r_{0}=10^5m$, $\beta=10^{-3}/s$ and $\chi=-4/3$, then the constant velocity is $v\simeq0.00074c\simeq221359m/s$ and the total travel time is $\Delta t=1.01445s$. Also considering the accelerated motion with $v=r_{0}R\nu_{B}$, together with Eq.\eqref{tsolution2}, we find 
\begin{equation}
\Delta t(\chi)=\frac{1}{\beta}\log\Bigg(\frac{\nu_{B}+\beta\mathcal{G}(\sqrt[3]{\frac{\chi}{1+\chi}},\chi)}{\nu_{B}-\beta\mathcal{G}(\sqrt[3]{\frac{\chi}{1+\chi},\chi})}\Bigg),
\end{equation}
which thus requires that $\nu_{B}>\beta\mathcal{G}(k_{max},\chi)$. The result shows that the traveler will spent $\Delta t=1.01394s$ to traverse the wormhole and he will stop his trip with a velocity $v_{f}\simeq221584m/s$. Therefore for nonrelativistic motion, this is a reasonable answer.

\section{conclusions}
\label{sec:conclusion}

In this paper we have discussed the traversability of the S-AdS wormholes. This kind of wormholes given in Eqs.\eqref{Wmetric} and \eqref{solution} does not tend to flat spacetime far away from the throat, and hence they, as the interior regions, need to match an exterior vacuum at the maximum radius $r_{max}$ of the wormholes where the shape functions vanish $b(r_{max})=0$. We studied the NEC of the matter content that supports the wormhole, the tidal force that the traveler going through the wormhole would have felt, the total time that it takes to traverse the wormhole, which consist of the criteria of the traversability. Only for the static cases, we investigated the geodesics in the plane $\theta=\pi/2$.

We considered the S-AdS wormholes in the framework of $f(T)$ gravities. $f(T)$ theories are direct extensions of the Teleparallel Equivalent of General Relativity, and have made considerable achievements in various aspects of the gravitation researches. For our works, the particular choice of proper tetrad\cite{Lin2019} given in Eq.\eqref{tetrad} is of special interest because it allows one to admit the classical solutions of General Relativity into the $f(T)$ gravities. Therefore the S-AdS wormholes that can only be supported by exotic matter in General Relativity now can be supported by the modifications of the Teleparallel Equivalent of General Relativity, allowing the actual matter to satisfy the NEC.

We mainly considered the static case and two specific evolving cases of the S-AdS wormholes. Particularly, We have described the geodesic's behavior in the plane $\theta=\pi/2$ of static wormholes. The test particle may radially go through the wormhole. However, if the initial angular velocity is not zero, the test particle may be reflected by the wormhole at a reversal point. Besides this, by demanding that the tidal force felt by the traveler should not exceed the Earth's gravity, we obtain the upper bound of the traveler's velocity in each case. This gives rise to the concern of the travel time\cite{Cataldo:2017ard,Morris1988Wormholes}. Since the wormholes have limited sizes $r_{0}\le r\le r_{max}$, we have in fact considered only the time it takes to through this range of radial distance. The trip beyond this part may in fact be in an ordinary vacuum spacetime like Schwarzschild one. To estimate the traverse time, we take some reasonable assumptions to make our calculations simple and valid.

The present work may be potentially extended to several directions, since we have not paid much attention to the exterior spacetime that matches to the interior wormhole. One can consider a specific vacuum and its junction conditions at $r_{max}$, the travel time and tidal force in the exterior vacuum then may provide constraints on the size of the interior wormhole and the form of $f(T)$. Moreover, the non-radial travels are also of physical interests in which they may provide more insights of wormholes.

\section*{Acknowledgement}

This work is supported by the National Science Foundation of China under Grant No.12105179.	

%

\nocite{*}

\end{document}